\documentstyle[12pt,prd,aps]{revtex}
\def\be{\begin{equation}}
\def\en{\end{equation}}
\def\bea{\begin{eqnarray}}
\def\ena{\end{eqnarray}}

\newcommand{\sgn}{\mbox{sgn}}
\begin{document}
\title{GRAVITATIONAL RADIATION FROM MONOPOLES \\ CONNECTED BY STRINGS} 
\author{Xavier Martin and Alexander Vilenkin}
\address{Institute of Cosmology, Department of Physics and Astronomy,
Tufts University, Medford, MA 02155, USA}
\maketitle

\begin{abstract}
Monopole-antimonopole pairs connected by strings can be formed as
topological defects in a sequence of cosmological phase
transitions. Such hybrid defects typically decay early in the history
of the universe but can still generate an observable background of
gravitational waves. We study the spectrum of gravitational radiation
from these objects both analytically and numerically, concentrating on
the simplest case of an oscillating pair connected by a straight
string.
\end{abstract}

\section{Introduction}
Monopoles connected by strings can be formed in a sequence of symmetry
breaking phase transitions in the early universe
\cite{kibble,shelvil}. The simplest sequence of this sort is \be
G\rightarrow H\times U(1)\rightarrow H.\label{symbreak} \en For a
semi-simple group $G$, the first of these phase transitions gives rise
to monopoles which get connected by strings at the second phase
transition. If both of these phase transitions occur during the
radiation era, then the average monopole separation is always smaller
than the Hubble radius, and when monopole-antimonopole (M\={M}) pairs
get connected by strings and begin oscillating, they typically
dissipate the bulk of their energy to friction in less than a Hubble
time \cite{shelvil,hkr}.

A more interesting possibility arises in the context of inflationary
scenario \cite{vilrev}, when monopoles are formed during inflation but
are not completely inflated away. Strings can either be formed later
during inflation, or in the post-inflationary epoch. In this case, the
strings connecting M\={M} pairs can be very long. The correlation
length of strings, $\xi$, can initially be much smaller than the
average monopole separation, $d$; then the strings connecting
monopoles have Brownian shapes. But in the course of the evolution,
$\xi$ grows faster than $d$, due to small loop production, and to the
damping force acting on the strings. Eventually, $\xi$ becomes
comparable to the monopole separation, and we are left with M\={M}
pairs connected by more or less straight strings. At later times, the
pairs oscillate and gradually lose their energy by gravitational
radiation and by radiation of light gauge bosons (if the monopoles
have unconfined magnetic charges). When the energy of a string
connecting a pair is dissipated, the monopole and antimonopole
annihilate into relativistic particles.

The gravitational waves emitted by oscillating M\={M} pairs add up to
a stochastic background which can have an observable intensity
\cite{vilrev}. In order to calculate the spectrum of this background,
one first needs to find the radiation spectrum produced by an
individual oscillating pair. This is our main goal in the present
paper.

In contrast to the case of cosmic string loops, the dynamics of M\={M}
pairs connected by strings has not been studied in any detail. We
shall therefore concentrate on the simplest case of an oscillating
pair connected by a straight string, for which the equations of motion
can be solved exactly. Apart from its simplicity, this system has the
advantage of being close to monopole-string configurations one would
expect to find in the early universe.

After reviewing the dynamics of monopoles connected by strings in the
next Section, we calculate the gravitational radiation spectrum from
an oscillating pair in Section \ref{gravsec}, then sum up and discuss
our results in Section \ref{concl}. Some technical details are given in
the Appendices.

\section{Equations of motion}
The characteristic monopole radius $\delta_m$ and string thickness
$\delta_s$ are determined primarily by the corresponding symmetry
breaking scales, $\eta_m$ and $\eta_s$. Typically, $\delta_m \sim
\eta_m^{-1}$ and $\delta_s \sim \eta_s^{-1}$. In order to have
monopoles connected by long strings, the two symmetry breaking scales
should be well separated, $\eta_s \ll \eta_m$, and thus the monopole
radius is much smaller than the string thickness, $\delta_m \ll
\delta_s$.

Assuming that the string length is much greater than its thickness, we
can treat monopoles as point particles and strings as infinitely thin
lines. The dynamics of a M\={M} pair connected by a string can then be
described by the action \be
I=-m\int ds_1-m\int ds_2 -\mu \int dS.\label{action} \en
Here, $\mu$ is the string tension (which is equal to the mass of
string per unit length), the first two integrations are over the
monopole and antimonopole worldlines and the third is over the string
worldsheet.

The last term in Eq. (\ref{action}) is the Goto-Nambu action for the
string. Its variation gives the standard string equations of motion
\be \delta I_{string} =\mu \int \partial_a (\sqrt{-\gamma}\gamma^{ab}
x^\mu_{,b}) \delta x_\mu d\zeta_0 d\zeta_1 -\mu \int \partial_a
(\sqrt{-\gamma}\gamma^{ab} x^\mu_{,b} \delta x_\mu ) d\zeta_0 d\zeta_1
\label{stract} \en
where $a=0,1$, $\zeta_a$ are a set of internal coordinates for the
string worldsheet, $x^\mu_{,a}=\partial_a (x^\mu )$, and $\gamma_{ab}
=g_{\mu \nu} x^\mu_{,a} x^\nu_{,b}$ is the two dimensional worldsheet
metric. The first term gives the equations of motion for the string
which take a particularly simple form in the transverse traceless
gauge \be
\dot{\bf x}^2 +{{\bf x}'}^2 =1 \ ,\ \dot{\bf x}\cdot {\bf x}'=0,
\label{gauge} \en
where the set of internal coordinates was taken to be $(t,\sigma )$,
and primes and dots refer to derivatives with respect to $\sigma$ and
$t$, respectively. In this gauge, the dynamical equations are simply
\be \ddot{\bf x}={\bf x}'' \label{streq} \en
and can be solved exactly as \be
{\bf x}(t,\sigma )=\frac{1}{2} [ {\bf a}(\sigma -t)+{\bf b}(\sigma +t)] /2,
\label{stsol} \en
with the gauge conditions (\ref{gauge}) taking the form \be
{{\bf a}'}^2 ={{\bf b}'}^2 =1. \label{constreq} \en
Since the string has two end points (one at each monopole), the
spatial parameter must be restricted to an interval $[\sigma_1
(t),\sigma_2 (t)]$ where \be
{\bf x}_i (t)={\bf x}(t,\sigma_i (t))\en
are the positions of the monopoles.

The second integral in Eq. (\ref{stract}) can be turned into a boundary
term and thus contributes to the variation of the monopole and
antimonopole worldlines \be
\int \partial_a (\sqrt{-\gamma}\gamma^{ab} x^\mu_{,b} \delta x_\mu )
d\zeta_0 d\zeta_1 =-\int \lambda_{1a}\gamma^{ab}x^\mu_{,b}\delta
x_{1\mu} ds_1 -\int \lambda_{2a}\gamma^{ab}x^\mu_{,b}\delta x_{2\mu}
ds_2 ,\en
where $\lambda_{1a}$ and $\lambda_{2a}$ are unit vectors orthogonal to
their respective worldlines and oriented into the string worldsheet,
and $x^\mu_{,b}$ are evaluated on the monopoles worldline at
$(t,\sigma_i (t))$. By definition, the same vector expressed in
external coordinates is \be
\lambda_i^\mu (t)=\lambda_{ia} \gamma^{ab}x^\mu_{,b}=\pm \gamma_i (t)
[\sigma_i '(t)\dot{x}^\mu (t,\sigma_i (t))+{x'}^\mu (t,\sigma_i (t))],\en
where $\gamma_i =(1-\dot{\bf x}_i^2 )^{-1/2}$ is the Lorentz factor of
the monopoles. If we now add the terms coming from the variation of
the monopole and antimonopole actions, we get the equations of motion
for them in the form \cite{carter} \be
\frac{d^2 x_i^\nu}{ds^2}=\frac{\mu}{m} \lambda_i^\nu ,\label{moneq} \en
where $\mu$ is the mass per unit length (and tension) of the
string and $m$ is the monopole mass. Since $\lambda_i^\mu$
are unit vectors, it follows from (\ref{moneq}) that \be
a=\mu /m,\label{defa} \en
is the proper acceleration of the monopoles. By multiplying equations
(\ref{moneq}) by $\dot{x}_{i\nu}$, it can be seen that only three of
the four equations are independent. The time component equation of the
system (\ref{moneq}), expressing the exchange of energy between the
length of string created or destroyed at the monopole end, and the
kinetic energy of the monopole, takes the very simple integrable form
\be a\dot{\sigma}_i =\pm \dot{\gamma}_i ,\en
while the spatial equations can be put in the form \be
\gamma_i^3 \ddot{\bf x}_i (t)=\pm a\left( \frac{{\bf x}'}{({\bf x}')^2}
\right) (t,\sigma_i (t)) =a\gamma_i^{(s)} {\bf n}_i .\label{moneq2} \en
Here, $\gamma_i^{(s)}=|{\bf x}'|^{-1}$ is the Lorentz factor of the
string at the location of the monopole, and ${\bf n}_i$ is a unit
vector pointing from the monopole in the direction of the string.

The complete set of dynamical equations for the system of two
monopoles connected by a string is thus given by the systems of
equations (\ref{streq}) and (\ref{moneq2}) with the constraint
(\ref{gauge}). In general, the solutions of these equations are not
periodic.

Though the string part (\ref{gauge}-\ref{streq}) can be solved
immediately in the standard form (\ref{stsol}-\ref{constreq}), the
motion of the monopoles is impossible to solve in general because of
the presence of $\sigma_i (t)$, which is one of the unknowns, as a
parameter of ${\bf x}'$ in equation (\ref{moneq2}). However, two sets
of particular exact solutions can be found by assuming either that
$\sigma (t)$ is constant or that ${\bf x}'$ does not depend on its
spatial parameter, thereby removing the problem.

In the first solution, the string has the form of a rotating rod of
length $l$ with the centrifugal force acting on the monopoles balanced
by the string tension: \begin{mathletters} \label{rodsol} \bea
{\bf x}(t,\sigma ) & = & R\sin (\sigma /R) {\bf e} (t/R),\\ 
{\bf x}_i (t) & = & \pm (l/2){\bf e} (t/R),\\
\sigma_i (t) & = & \pm R\arcsin (l/2R),\ena \end{mathletters}
where ${\bf e} (\theta )=(\cos \theta ,\sin \theta )$ is the
radial unit vector associated with the angle $\theta$ and $l=((1+4a^2
R^2)^{1/2}-1)/a$. We note that in the case when the monopoles are
relativistic, that is when $aR\gg 1$, the energy of the string is much
larger than the energy of the monopoles, $E_s /E_m \sim aR$. We will
make use of this property to approximate the gravitational radiation
spectrum of this solution by that of a simple cosmic string loop
(rotating double line).

The second solution describes an oscillating pair of monopoles
connected by a straight string: \begin{mathletters} \label{strsol} 
\bea {\bf x}(t,\sigma ) & = & \sigma {\bf e},\\ 
{\bf x}_i (t) & = & \pm \frac{\sgn (t)}{a} (\gamma_0 -\sqrt{1+
(\gamma_0 v_0 -a|t|)^2}) {\bf e} \\
\sigma_i (t) & = & x_i (t) .\ena \end{mathletters}
Here $\bf e$ is a unit vector along the string, which we choose to be
directed along the $x$-axis, $a$ is the monopole proper acceleration
defined in (\ref{defa}), $v_0$ and $\gamma_0 =(1-v_0^2)^{-1/2}$ are
respectively the maximum velocity and Lorentz factor of the monopoles,
reached at $t=0$. The monopole and antimonopole meet at $t=0$ and
could be expected to annihilate. However, this solution is considered
as an approximation for an almost straight configuration where the
monopole and antimonopole would merely come close to each other and
would not collide. Besides, as we already mentioned, the monopole
radii are much smaller than the string thickness, thus the
monopoles are not likely to collide, even for a straight string. A
peculiar feature of the solution (\ref{strsol}) is that the monopole
accelerations abruptly change direction when the monopoles meet and
pass one another.

The solution (\ref{strsol}) is valid for $|t|\leq \gamma_0 v_0 /a$. At
$t=-\gamma_0 v_0 /a$, the monopoles are at rest, with the string
having its maximum length, $L=2(\gamma_0 -1)/a$. At $t=+\gamma_0 v_0
/a$, the monopoles come to rest again, with their positions
interchanged. As far as gravitational effects are concerned, since the
monopole and antimonopole have the same mass, they are identical, and the
full period of motion is $T=2\gamma_0 v_0 /a$. On the other hand, if
electromagnetic effects are considered, the monopole and antimonopole
have opposite charges, and thus equation (\ref{strsol}) describes only
half a period, $T=4\gamma_0 v_0 /a$, the other half period being
obtained by exchanging the positions of the monopole and antimonopole:
\be {\bf x}_1 (t+T/2)={\bf x}_2 (t)=-{\bf x}_1 (t).\en

In the following section, we shall study the gravitational radiation
from an oscillating pair described by the solution
(\ref{strsol}). Radiation from the rotating rod configuration will be
discussed in Appendix \ref{specrod}.

\section{Gravitational radiation} \label{gravsec}
The power in gravitational radiation from a weak, isolated, periodic
source to lowest order in $G$, can be found from the following
equations, without any further assumptions about the source
\cite{wein} \begin{mathletters} \label{gravspec} \be
P=\sum_n P_n = \sum_n \int d\Omega \frac{dP_n}{d\Omega}, \en \be
\frac{dP_n}{d\Omega} = \frac{G\omega_n^2}{\pi} (T_{\mu \nu}^*
(\omega_n ,{\bf k}) T^{\mu \nu} (\omega_n ,{\bf k})-\frac{1}{2}
|{T_\mu}^\mu (\omega_n ,{\bf k})|^2 ). \label{gravspecb} \en \end{mathletters}
Here, $dP_n /d\Omega$ is the radiation power at frequency $\omega_n =2
\pi n/T$ per unit solid angle in the direction of $\bf k$, $|{\bf k}|=
\omega_n$, $T$ is the period of the oscillation, and \be
T^{\mu \nu} (\omega_n ,{\bf k})=\frac{1}{T} \int_0^T dt \exp (i\omega
_n t)\int d^3 x \exp (-i{\bf k}\cdot{\bf x}) T^{\mu \nu} ({\bf x},t)
\label{fourtr} \en is the Fourier transform of the energy-momentum tensor.

For the solution (\ref{strsol}) considered in this section, since the
system is one dimensional, the energy-momentum tensor has only three
non zero components: $T^{00}$, $T^{01}$ and $T^{11}$. It also
satisfies the conservation equations $\nabla_\mu T^{\mu \nu} =0$ which
in Fourier space can be written simply \be
\omega_n T^{0\nu} =k^i T^{i\nu}.\label{conseq} \en
This means that $T^{\mu \nu}$ has in fact only one independent
component; the simplest choice is \be
T^{01} (t,{\bf x})=m(\gamma_0 v_0 -a|t|) [\delta ({\bf x}-x_1 (t) {\bf
e}) -\delta ({\bf x}+x_1 (t) {\bf e})], \en
which has no contribution from the string part of the system. Its
Fourier transform as defined by (\ref{fourtr}) can be simplified to
\begin{mathletters} \label{fft} \bea 
T^{01}(\omega_n ,{\bf k}) & = & m\gamma_0 v_0 I_n (u),\label{ffta} \\
I_n (u) & = & \int_0 ^1 \xi d\xi [\cos (n\pi (1-\xi -\frac{u}{v_0}+u
\sqrt{\xi^2 +1/(\gamma_0 v_0 )^2})) \label{fftb}\\
& & -\cos (n\pi (1-\xi +\frac{u}{v_0}-u\sqrt{\xi^2 +1/(\gamma_0 v_0
)^2}))], \nonumber \ena \end{mathletters}
where $\omega_n =n\pi a/(\gamma_0 v_0 )$ is the angular frequency of
the $n$-th mode and we have introduced the notation $u=k_x /\omega_n$.
The two other non-zero components of the Fourier transform of the
energy-momentum tensor can then be deduced from the conservation
equations (\ref{conseq}) as \begin{mathletters} \bea
T^{00}(\omega_n ,{\bf k}) & = & uT^{01}(\omega_n ,{\bf k}),\\
T^{11}(\omega_n ,{\bf k}) & = & \frac{1}{u}T^{01}(\omega_n ,{\bf k}).
\ena \end{mathletters}
The gravitational energy radiated in the mode $n$ can then be expressed
from (\ref{gravspecb}) as \be
P_n = 2G(n\pi \mu )^2 \int_0^1 du (\frac{1}{u}-u)^2
|I_n (u)|^2 ,\label{pwrs} \en
where $I_n (u)$ is given by (\ref{ffta}) and depends only on $u=k_x
/\omega_n$. This power spectrum can not be integrated in a closed
form. However, it is possible to get analytic approximations
at low and high frequency.

At high frequency, an expansion of (\ref{fft}) in $1/n$ can be made
as shown in Appendix \ref{expans}. First, a change of variable \be
\zeta =\pm \xi -u\sqrt{\xi^2 +1/(\gamma_0 v_0 )^2} \en
enables us to rewrite (\ref{fftb}) in the standard form (\ref{genint})
\begin{mathletters} \label{grint} \bea
I_n (u) & = & \frac{u}{(1-u^2 )^2} \int_{-1-\frac{u}{v_0}}^{1-\frac{u
}{v_0}} f(u,\zeta )\cos [n\pi (1+\frac{u}{v_0}+\zeta ) ] d\zeta ,
\label{grinta} \\
f(u,\zeta ) & = & 2(\zeta^2+\frac{1-u^2}{\gamma_0^2v_0^2})^{1/2}-
\frac{1-u^2}{\gamma_0^2 v_0^2} (\zeta^2 +\frac{1-u^2}{\gamma_0
^2 v_0^2})^{-1/2} . \ena \end{mathletters}
The expansion is obtained by repeatedly integrating by parts in Eq. 
(\ref{grinta}). The dominant term is \be
I_n (u) \approx \frac{uv_0}{(1-u^2 )}\left( \frac{4}{1-u^2 v_0^2}-\frac{1-
v_0^2}{(1+uv_0)^3} -\frac{1-v_0^2}{(1-uv_0)^3} \right) .\en
To get the interval of validity of this expansion, this term must be
compared with the next as shown in Eq. (\ref{valid}). For simplicity
and because it is the most interesting case, we shall assume in the
following that the monopoles are ultra-relativistic, $\gamma_0 \gg 1$.
Then, gravitational radiation is beamed in the monopole's direction of
motion, into a cone of a small opening angle $\theta \sim
1/\gamma_0$. Thus, the main contribution to the total power (\ref{gravspec})
comes from values around $u\simeq 1$ and the comparison of the two
terms can be performed there. This gives \begin{mathletters} \bea
[f']_{-1-\frac{u}{v_0}}^{1-\frac{u}{v_0}} & \approx & (1-u^2)\gamma_0^4 ,\\
\left[ f'''\right] _{-1-\frac{u}{v_0}}^{1-\frac{u}{v_0}} & \approx &
(1-u^2)\gamma_0^8 , \ena \end{mathletters}
so that the expansion is valid as long as \be n\gg \gamma_0^2 .\en

The dominant term of the power spectrum can be expressed as \be
P_n \approx \frac{2G\mu^2}{n^2\pi^2v_0^3} \int_{1-v_0}^1 dw \left[
\frac{4v_0^2}{w(2-w)}+\frac{\gamma_0^2 w-1}{\gamma_0^4 w^3}+
\frac{1+v_0^2-w}{\gamma_0^2 (2-w)^3} \right] ^2 .\label{asex} \en
The $1/n^2$ behavior at infinity could be expected since the second
derivative of the stress energy tensor has a discontinuity at
$t=0$. When $\gamma_0$ is also assumed to be large, that is the
monopoles reach ultrarelativistic velocities, the integral (\ref{asex})
can be simplified to \be
P_n \approx \frac{64}{5\pi^2} \left( \frac{\gamma_0}{n} \right) ^2 G
\mu^2. \label{hnb} \en

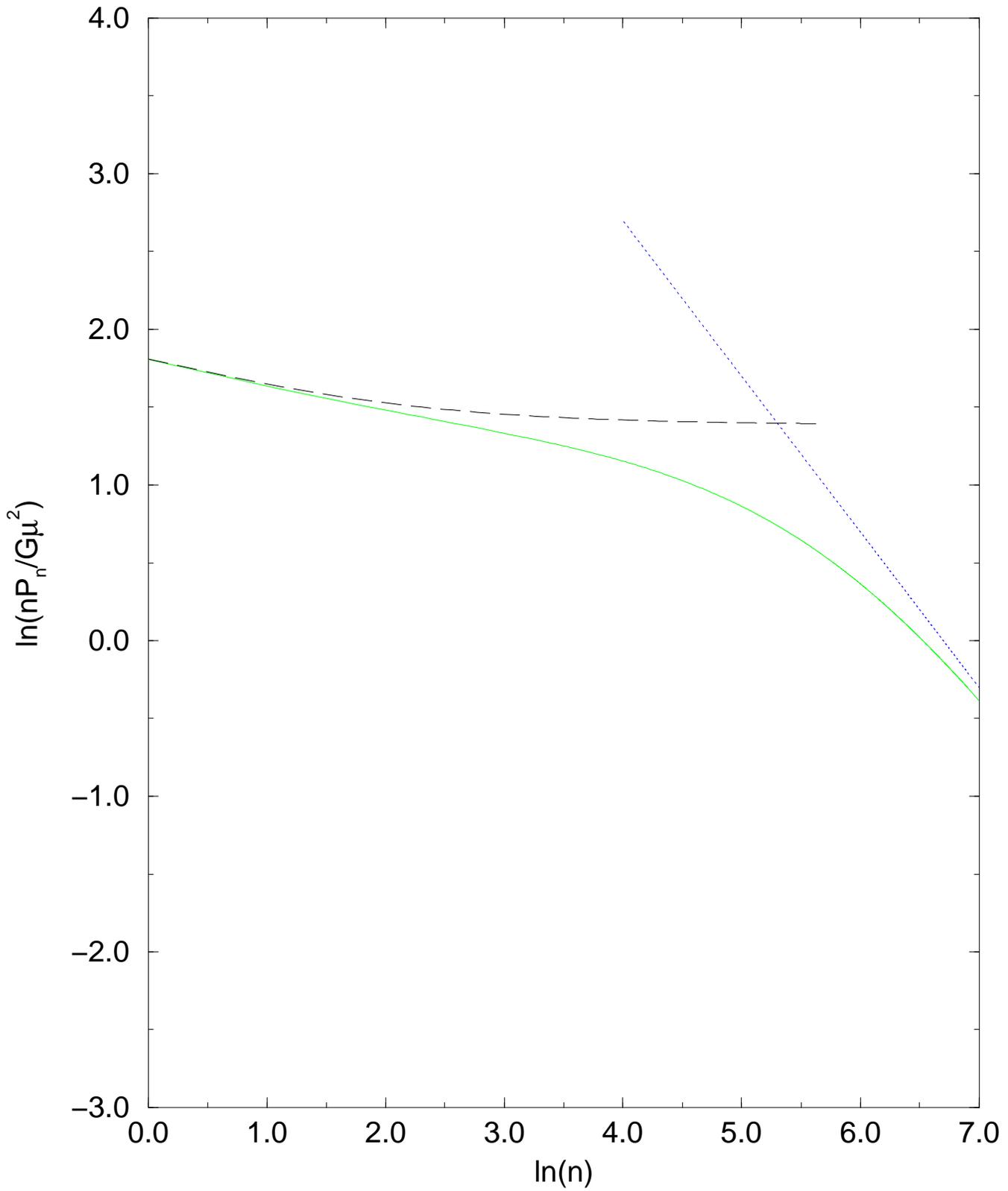
\begin{figure}
\caption{A log-log plot of the gravitational radiation
spectrum $nP_n /(G\mu^2 )$ in the case $\gamma_0 =25$ (solid line)
with its high frequency approximation (\protect\ref{hnb}) (dashed line)
and low frequency approximation (\protect\ref{lnb}) (dotted line).}
\label{ps25}
\end{figure}

For $n\ll \gamma_0^2$ with $\gamma_0 \gg 1$, it is legitimate to
neglect the terms $(\gamma_0 v_0 )^{-2}$ in the square roots of
equation (\ref{fftb}). This gives \be
T^{01} (\omega_n,{\bf k})\approx 8m\gamma_0 u\left( \frac{\sin (n\pi
(1-u)/2)}{n\pi(1-u^2)}\right) ^2 ,\en
and the power in the $n$-th mode can be simplified to
\be P_n \approx \frac{16}{n\pi} G\mu ^2 \int_0^{n\pi /2} du \frac{\sin^4
u}{u^2 (1-\frac{u}{n\pi})^2}. \label{lnb} \en 
It is interesting to note that this low frequency behavior is {\em
independent} of the maximum Lorentz factor of the monopole $\gamma_0$,
as long as it remains large. Furthermore, for large $n$, it is
possible to neglect $u/n\pi$ and extend the integration to infinity so
that we get \be
P_n \approx \frac{16}{n\pi} G\mu ^2 \int_0^{\infty} du \frac{\sin^4
u}{u^2}=\frac{4G\mu^2}{n}.\en
The low frequency behavior (\ref{lnb}) has been plotted in Figure
\ref{ps25} and is indeed well approximated by $4G\mu^2 /n$ for
$n\gtrsim 30$. From a cosmological point of view, the quantity of
interest is the gravitational energy per logarithmic interval $nP_n$
which is therefore quasi-constant in the frequency interval
$\gamma_0^2 \gg n\gtrsim 30$. Curiously, the low frequency spectrum
for the other solution, the rotating rod (\ref{rodsol}), though it is
very different from the straight string solution, also behaves like
$1/n$ (see Appendix \ref{specrod}).

The power spectrum can be computed numerically by integrating
successively (\ref{fft}) and (\ref{pwrs}). The evaluation was done for
various values of $\gamma_0$ and mostly exhibits a smooth evolution
from the low frequency (\ref{lnb}) to the high frequency (\ref{hnb})
behaviors. An example with $\gamma_0=25$ is shown in Figure
\ref{ps25}.

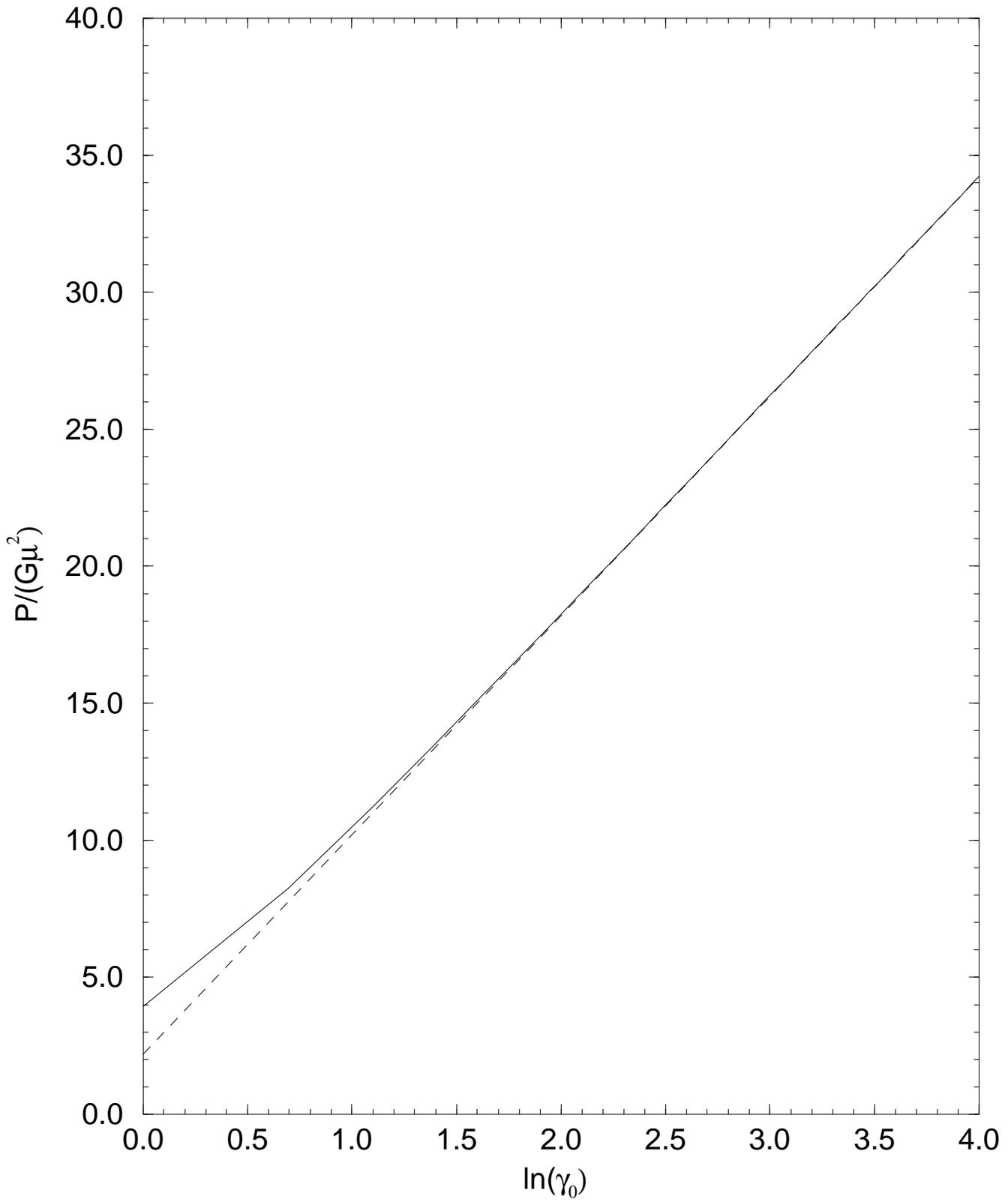
\begin{figure}
\caption{Total gravitational power emitted (solid line) and its
empirical algebraic approximation (\protect\ref{totpowa}) (dashed
line) as functions of $\ln (\gamma_0 )$.}
\label{totpowf}
\end{figure}

It is also useful to know the behavior of the total gravitational
energy loss rate $P$ as a function of $\gamma_0$. Once again only a
numerical solution is possible. However, instead of adding up the
power in different modes, it is much faster to compute $P$ directly
without going to Fourier space. The calculation is outlined in
Appendix \ref{dirtotpow}, and the resulting radiation rate is plotted
in Figure \ref{totpowf} as a function of $\gamma_0$. Empirically, it
can be closely approximated by \be 
P(\gamma_0)\simeq (8\ln (\gamma_0)+2.2)G\mu^2 ,\label{totpowa} \en
which is consistent with a spectrum behaving first like $4G\mu^2/n$ up
to $n\sim \gamma_0^2$ and then like $(64/5\pi^2)(\gamma_0/n)^2G \mu
^2$. For large values of $\gamma_0$, the flat low frequency part of
the spectrum makes the dominant contribution to the total
gravitational power emitted. In the case $\gamma_0 =25$ considered
above, the power computed numerically is $P=28.0G\mu^2$ (in agreement
with a summation of all the modes of the spectrum found in Figure
\ref{ps25}) while the algebraic approximation gives $P\simeq
28.3G\mu^2$ with a relative error of only $1\%$.

\section{Conclusions} \label{concl}

We have analyzed the gravitational radiation from an oscillating
monopole-antimonopole pair connected by a straight string. The motion
of the pair is described by Eq. (\ref{strsol}). The gravitational
radiation is emitted at a discrete set of frequencies $\omega_n
=n\omega_1$, where $\omega_1 =\pi a/\gamma_0 v_0$, $a=\mu /m$ is the
proper acceleration of the monopoles, $\mu$ is the string tension, $m$
is the monopole mass, $v_0$ is the highest velocity reached by the
monopoles, and $\gamma_0 =(1-v_0^2)^{-1/2}$ is the corresponding
Lorentz factor. In the most interesting case of ultrarelativistic
motion, $\gamma_0 \gg 1$, most of the radiation is emitted in the
range $1\lesssim n\lesssim \gamma_0^2$, with a spectrum $P_n\approx
4G\mu^2/n$. For $n\gg 1$, the spectrum can be approximated as
continuous with \be
dP/d\omega \approx 4G\mu^2 /\omega .\en
At higher frequencies, $dP/d\omega \propto \omega^{-2}$. The total
radiation power is \be
P\approx 8G\mu^2 \ln (\gamma_0 ).\en

The one dimensional solution (\ref{strsol}) should be regarded as an
approximation for a more general configuration of monopoles connected
by a nearly straight string. In a more realistic case, the $1/\omega$
and $1/\omega^2$ behavior is expected to be modified for sufficiently
large $n\gg n_c$.The characteristic value $n_c$ and the corresponding
frequency $\omega_c$ can be estimated from $\omega_c \sim (\Delta
l)^{-1}$, $n_c \sim l/\Delta l$, where $\Delta l$ is the monopole
separation at which deviations from the straight-line shape become
important and $l$ is the maximum extent of the string. Typically,
$\Delta l$ is comparable to the minimum distance between the monopoles
as they pass one another.

We cannot tell from our analysis how the spectrum is modified at
$\omega >\omega_c$. This remains a problem for future research. We
expect that eventually $P_n$ will fall off exponentially at
$n\rightarrow \infty$, but there may also be some intermediate
regime. Since no solutions of the equations of motion are known in
which the string would deviate from a straight-line shape, this
problem will probably have to be tackled numerically. In particular,
one could employ the numerical simulations of the monopole-string
system that are currently being developped \cite{xavier}.

Though in the simplest models of symmetry breaking all the magnetic
flux of the monopoles is confined in the strings, in more realistic
models, stable monopoles can have unconfined magnetic charges. In such
a case, monopoles can lose energy by radiating gauge quanta. The gauge
fields associated with the magnetic charge may include electromagnetic
or gluon fields, but may also correspond to broken gauge symmetries
and have non zero masses. The gauge boson radiation is important since
it can greatly affect the lifetime of the pair and thus the total
output of gravitational waves. An evaluation of the radiation of
massless or massive gauge bosons by monopoles will be given elsewhere
\cite{abx}.

\section*{Acknowledgements}
This work was supported in part by the National Science Foundation.

\appendix
\section{Gravitational radiation spectrum for a ``rotating rod'' solution}
\label{specrod}
The rotating rod solution (\ref{rodsol}) does not seem likely to arise
naturally in the early universe but could give some insight into what
happens for a configuration very different from the one-dimensional
solution (\ref{strsol}). We shall concentrate on the case when the
monopoles are relativistic, that is when $aR\gg 1$. In that case, it
is easy to check that the energy of the string is much larger than the
energy of the monopoles, $E_s /E_m \sim aR\gg 1$. This means that the
monopoles can be ignored in calculating the gravitational radiation of
the system. As for the string part of the system, it should be well
approximated by the straight rotating double line solution \be
{\bf x}(t,\sigma )=R\sin (\frac{\sigma}{R}) {\bf e} (t/R) ,\label{loopsol}
\en with $|\sigma |\leq \pi R$, as long as the wavelengths of the
gravitational waves remain large compared to the length difference
of the rotating string segments in the solution (\ref{rodsol}) and its
approximation (\ref{loopsol}): $n\ll aR$. Since the loop (\ref{loopsol})
is made of two straight strings, its mass per unit length $\tilde\mu$
should be half that of the straight rod solution: \be
\tilde\mu =\mu /2.\label{tildemu} \en 
The straight loop solution (\ref{loopsol}) is singular at its ``end
points'' $\sigma =\pm \pi R/2$, which move at the speed of light. This
singularity results in a $1/n$ decay of its gravitational radiation
spectrum and thus in a divergent total power. This is not really a
problem since our approximation is only valid at low frequencies,
$n\ll aR$. However, it is for this reason that, though the
gravitational radiation of similar loop solutions has been studied in
the past \cite{vv,burden}, the calculation for this particular solution
do not appear to have ever been done.

Using the standard formula for gravitational radiation from a
string in the direction of the unit vector $\bf k$, \cite{garf} \bea
\frac{dP_n}{d\Omega} ({\bf k})=8\pi G\tilde\mu ^2 n^2 & & \left\{ |I_n
({\bf n}_1 ) J_n ({\bf n}_1 )-I_n ({\bf n}_2 ) J_n ({\bf n}_2 )|^2
+\right. \\
& & \left. |I_n ({\bf n}_1 ) J_n ({\bf n}_2 )+I_n ({\bf n}_2 ) J_n
({\bf n}_1 )|^2 \right\} ,\nonumber \ena
where ${\bf n}_1$ and ${\bf n}_2$ are unit vectors orthogonal to ${\bf
k}$ and to one another, $I_n ({\bf n})$ and $J_n ({\bf n})$ are
defined as \begin{mathletters} \bea
I_n ({\bf n}) & = & \frac{1}{L} \int_0^L d\zeta {\bf a}'(\zeta )\cdot
{\bf n} \exp [-\frac{in\omega}{2}(\zeta +{\bf k}\cdot{\bf a}(\zeta ))],\\
J_n ({\bf n}) & = & \frac{1}{L} \int_0^L d\zeta {\bf b}'(\zeta )\cdot
{\bf n} \exp [\frac{in\omega}{2}(\zeta -{\bf k}\cdot{\bf b}(\zeta
))],\ena \end{mathletters}
${\bf a}(\zeta )$ and ${\bf b}(\zeta )$ are the generators of the
string worldsheet as defined in equation (\ref{stsol}), $L$ is their
period and $\omega$ is the angular frequency of the loop. For the loop
solution (\ref{loopsol}), after integrating the gravitational power
over all directions and replacing $\tilde\mu$ by its value
(\ref{tildemu}), we have \bea
P_n & = & 8\pi^2 G\mu^2 n^2 \int_0^{\pi/2} \sin x dx [\frac{\cos^4
x}{\sin^4 x}J_n^4 (n\sin x)+6\frac{\cos^2 x}{\sin^2 x} \label{loopspec} \\
& & J_n^2 (n\sin x){{J_n}'}^2 (n\sin x)+ J_n^4 (n\sin x){{J_n}'}^2 
(n\sin x)],\nonumber \ena
where $J_n (x)$ are Bessel functions of the first kind. This integral
can be computed numerically and gives a power spectrum which behaves
almost exactly like $5.75G\mu^2 /n$, even at low frequencies. The
spectrum of the rotating rod solution (\ref{rodsol}) should therefore
behave like $1/n$ for low-frequency modes $n\ll aR$, but is expected
to fall exponentially at high frequencies (since the stress energy
tensor of the system, and all its derivatives, are regular).

\section{Large-$\lowercase{\bf n}$ expansion} \label{expans}
In this Appendix, we develop a method for deriving large-$n$ asymptotic
expansions for integrals of the form \be
I_n=\int_a^{a+2} f(u)\cos (n\pi (u-a))du,\label{genint} \en 
where $f$ is a regular function which does not depend on n. Examples
of such expressions are found in this paper in equation (\ref{grint}),
and more generally in problems involving Fourier transforms. The
trick is to integrate by parts the cosinus \bea
I_n & = & \frac{1}{n\pi}[ f(u)\sin (n\pi (u-a))]_a^{a+2} -
\frac{1}{n\pi} \int_a^{a+2} f'(u)\sin (n\pi (u-a))du \nonumber \\ 
& = & \frac{1}{n^2\pi^2} [ f'(u)\cos (n\pi (u-a))]_a^{a+2} -
\int_a^{a+2} f''(u)\cos (n\pi (u-a))du.\ena 
The last integral is of the same general form as the initial
expression (\ref{genint}) so that this expansion can be iterated to
get \be 
I_n = \sum_{l=1}^{\infty} (f^{(2l-1)}(a+2)-f^{(2l-1)}(a))\frac{(-1)
^{l-1}}{(n\pi )^{2l}}.\en 
In this paper, we are only concerned with an expansion for large $n$
to the first significant order \be 
I_n =\frac{1}{n^2\pi^2} [f'(u)]_a^{a+2} +O(\frac{1}{(n\pi)^4}).\en
The interval of validity of this expansion can be evaluated by
requiring that the first neglected term in the expansion be much
smaller than the one we keep. This gives \be
n^2 \gg \frac{[f'''(u)]_a^{a+2}}{[f'(u)]_a^{a+2}}. \label{valid} \en

\section{Direct calculation of the total gravitational power}
\label{dirtotpow}
Here, we derive an expression for the total gravitational power
radiated by the one-dimensional solution (\ref{strsol}). The metric
perturbation $h^{\mu \nu}=g^{\mu \nu}-\eta^{\mu \nu}$ induced by a
weak source at a large distance $r$ in the direction of the unit
vector $\bf n$ can be expressed as \begin{mathletters} \bea
\bar{h}^{\mu \nu} (t,r{\bf n}) & = & \frac{4G}{r} \int T^{\mu \nu}
(t_r,{\bf x}') d^3x',\\
\bar{h}^{\mu \nu} & = & h^{\mu \nu}-\frac{1}{2}{h_\rho}^\rho \eta^{\mu
\nu}, \\ t & = & t_r +|{\bf x}'-{\bf x}(t_r )| ,\ena \end{mathletters}
where $\eta^{\mu \nu}$ is the Minkowski metric. The gravitational
power emitted by the source can then be expressed as \be
\frac{dP}{d\Omega} (t)=\frac{r^2}{32\pi G} (\bar{h}^{\mu \nu} 
\bar{h}_{\mu \nu} -\frac{1}{2} ({\bar{h}^\mu}_\mu )^2),\label{totpowg} \en 
which no longer depends on the distance to the source $r$. In the
present case, there are only three non zero terms $h^{\mu\nu}$ given
by \begin{mathletters} \label{metvar} \bea
h^{00} & = & h_0 +\frac{mn_x}{2} \left( \frac{du_{1r}^2}{du}+\frac{du_
{2r}^2 }{du} \right) ,\\
h^{01} & = & \frac{m}{2} \left( \frac{du_{1r}^2}{du}+\frac{du_{2r}
^2}{du} \right) ,\\
h^{11} & = & h_1 -\frac{mn_x}{2} \left( \frac{du_{1r}^2}{du}+\frac{du
_{2r}^2 }{du})+m(\frac{1+2u_{2r}^2}{\gamma_2 -n_x u_{2r}}+ \frac{1+2u
_{1r}^2 }{\gamma_1 +n_x u_{1r}} \right) ,\ena 
where $h_0$ and $h_1$ are constants which do not contribute to the total
gravitational power, $u_{1r}=\gamma_0 v_0 -a|t_{1r}|$ and
$u_{2r}=\gamma_0 v_0 -a|t_{2r}|$ are functions of the retarded time
for each monopole which can be expressed as functions of $u=\gamma_0 
v_0 -a|t|$ by
\bea u_{ir} & = & \frac{(-1)^i n_x \sqrt{v_i^2+1-n_x^2}-v_i}{1-n_x^2}, \\
v_i & = & (-1)^i n_x \gamma_0 -u.\ena
What we are really interested in is the gravitational radiation power
averaged over a period. This is obtained by integrating equation
(\ref{totpowg}) over all directions and then averaging over time. This
gives \be
P=\frac{r^2}{16G\gamma_0 v_0}\int_0^{\gamma_0 v_0}du\int_0^1 dn_x
(\dot{\bar{h}}^{00}+\dot{\bar{h}}^{11}+\dot{\bar{h}}^{01})(\dot{
\bar{h}}^{00}+\dot{\bar{h}}^{11}-\dot{\bar{h}}^{01}).\en \end{mathletters}
Though the expressions (\ref{metvar}) are complicated, finding the
average total emitted power $P(\gamma_0 )$ only requires a double
numerical integration. Since the evaluation of each point of the
spectrum also required a double integration, this method is much more
efficient than simply summing the power of all the modes.

\begin{thebibliography}{99}
\bibitem{kibble} T.W.B. Kibble, {\it J. of Phys.} {\bf A9}, 1387 (1976).
\bibitem{shelvil} For a review of topological defects in cosmology, see 
A. Vilenkin and E.P.S. Shellard, {\it Cosmic strings and other
topological defects} (Cambridge University Press 1994).
\bibitem{hkr} R. Holman, T.W.B. Kibble and S.-J. Rey, {\it Phys. Rev.
Lett.} {\bf 69}, 241 (1992).
\bibitem{vilrev} X. Martin and A. Vilenkin, {\it Phys.Rev. Lett.} {\bf
77}, 2879 (1996).
\bibitem{carter} B. Carter, Covariant mechanics of simple and
conducting strings and membranes, in {\it The formation and
evolution of Cosmic strings}, ed. by G. Gibbons, S. Hawking and
T. Vachaspati, pp 143-178 (Cambridge U.P. 1990).
\bibitem{wein} S. Weinberg {\it Gravitation and cosmology},
pp 260-266 (J. Wiley \& Sons N.Y. 1972).
\bibitem{ll} L.D. Landau and E.M. Lifschitz {\it The classical theory of
fields}'', pp 170-173 (Pergamon Press, London, 1971).
\bibitem{xavier} X. Martin and X. Siemens, work in progress.
\bibitem{abx} V. Berezinsky, X. Martin and A. Vilenkin paper in preparation.
\bibitem{vv} T. Vachaspati and A. Vilenkin, {\it Phys. Rev.} {\bf D31},
3052 (1985)
\bibitem{burden} C.J. Burden, {\it Phys. Lett.} {\bf 164B}, 277 (1985).
\bibitem{garf} D. Garfinkle and T. Vachaspati, {\it Phys. Rev.} {\bf 
D36}, 2229 (1987).
\end{thebibliography}
\end{document}